\documentclass[12pt]{iopart}
\usepackage{epsfig}
\usepackage{amssymb,bm}
\usepackage{graphicx}
\usepackage{epstopdf}
\usepackage{latexsym}
\usepackage{subfigure}
\usepackage[usenames,dvipsnames]{color}
\usepackage{hyperref}
\usepackage{moreverb}

\begin{document}

\title{Dipole-dipole interactions in optical lattices do not follow an inverse cube power law}

\author{M. L. Wall$^{1}$\footnote[3]{Corresponding author \quad E-mail: {\sf mwall.physics@gmail.com}} and L. D. Carr$^{1,2}$}
\address{$^1$Department of Physics, Colorado School of Mines, Golden, CO 80401, USA}
\address{$^2$Universit\"at Heidelberg, Physikalisches Institut, D-69120 Heidelberg, Germany}

\begin{abstract}
We study the effective dipole-dipole interactions in ultracold quantum gases on optical lattices as a function of asymmetry in confinement along the principal axes of the lattice.  In particular, we study the matrix elements of the dipole-dipole interaction in the basis of lowest band Wannier functions which serve as a set of low-energy states for many-body physics on the lattice.  We demonstrate that the effective interaction between dipoles in an optical lattice is non-algebraic in the inter-particle separation at short to medium distance on the lattice scale and has a long-range power-law tail, in contrast to the pure power-law behavior of the dipole-dipole interaction in free space.  The modifications to the free-space interaction can be sizable; we identify differences of up to 36\% from the free-space interaction at the nearest-neighbor distance in quasi-1D arrangements.  The interaction difference depends essentially on asymmetry in confinement, due to the $d$-wave anisotropy of the dipole-dipole interaction.  Our results do not depend on statistics, applying to both dipolar Bose-Einstein condensates and degenerate Fermi gases.  Using matrix product state simulations, we demonstrate that use of the correct lattice dipolar interaction leads to significant deviations from many-body predictions using the free-space interaction.  Our results are relevant to up and coming experiments with ultracold heteronuclear molecules, Rydberg atoms, and strongly magnetic atoms in optical lattices.

\end{abstract}

\maketitle

\section{Introduction}
Recent experimental progress in cooling heteronuclear polar molecules with large electric dipole moments~\cite{Carr_Demille_09,niKK2008, chotia_neyenhuis_11, Debatin_Takekoshi_11, Cho_McCarron_11, McCarron_Cho_11, Takekoshi_Debatin_12, Repp_Pires_13, Tung_Parker_13,Wu_Park_12,Heo_Wang_12}, Rydberg atoms~\cite{Anderson_Raithel_11}, and atoms with large magnetic dipole moments, in particular Chromium~\cite{griesmaier2005}, Erbium~\cite{Aikawa_Frisch_12}, and Dysprosium~\cite{Lu_Burdick_11,Lu_Burdick_12}, has sparked interest in the properties of ultracold dipolar gases.  While in many ultracold atomic systems interactions are short-range and well-modeled by a contact pseudopotential, the interactions in dipolar gases have a long-range and anisotropic character in free space, decaying as $1/r^3$ with the separation $r$ between particles.  These features of the dipole-dipole interaction have lead to a variety of intriguing theoretical proposals such as exotic pairing and bound states in ladder geometries~\cite{Dalmonte_Zoller_11,Wunsch_Zinner_11} and the realization of quantum liquid crystal states of matter~\cite{Quintanilla_Carr_09,Fregoso_Fradkin_09,Fregoso_Sun_09}.  Even for atoms with relatively weak dipole moments, such as Rubidium, dipole-dipole interactions can play a significant role~\cite{Vengalattore_Leslie_08,Vengalattore_Guzman_10}.  In this article, we show that the effective dipolar interaction in a lattice is not actually $1/r^3$ as commonly believed, but has a non-algebraic decay at moderate separations, and only behaves as $1/r^3$ for large separations.  Corrections of order 36-48\% arise for interactions at the nearest-neighbor distance in moderately confined quasi-low-dimensional scenarios.

A key component of our analysis is the presence of a continuous, periodic potential, which for ultracold atomic and molecular gases is provided by an optical lattice.  As first discovered by Kohn~\cite{Kohn_59}, Wannier functions, the most localized set of orthogonal single-particle states with the symmetries of the lattice, generally feature an exponential decay.  We find that the effective dipolar interaction in an optical lattice depends essentially on the exponential tails of the Wannier functions rather than only on their widths.  Hence, approximating the Wannier functions with localized functions which match only the mean width will fail to accurately capture the effective interaction in an optical lattice.

An additional essential ingredient for our findings is an asymmetry in the degree of confinement along the principal axes of the lattice due to the anisotropic character of the dipole-dipole interaction.  In this work, we characterize confinement using the curvature of a lattice site minimum.  Our work builds on a wealth of confinement-induced phenomena in ultracold quantum gases, such as confinement-induced resonances~\cite{Olshanii1998,Haller_Mark_10,Wall_Carr_12,Wall_Carr_13}, the fermionization of a 1D Bose gas~\cite{kinoshita2004}, and the Berezinskii-Kosterlitz-Thouless transition in a quasi-2D Bose gas~\cite{hadzibabic2006,Desbuquois_Chomaz_12}.  We stress, however, that our work does not deal with dipolar confinement-induced resonances such as those studied in Ref.~\cite{DipolarCE}, but rather on the modification of the effective interaction in an optical lattice due to the localization properties of the single-particle basis.  In dipolar gases, the effects of confinement have been studied in harmonic traps~\cite{Buechler_Demler_07,santos2000,santos2003}, and within the Gross-Pitaevskii approximation in a triple-well potential~\cite{Peter_Paw_12} and a 1D lattice~\cite{Wilson_Bohn_11}.  For the harmonic oscillator, it has been shown that strong confinement along the axis of a field orienting the dipoles leads to purely repulsive interactions in the weakly confined plane.  Additionally, because of the anisotropic character of the dipole-dipole interaction, the stability of a dipolar BEC displays a strong dependence on anisotropy in external confinement, and stable solutions can even take surprising forms such as the "red blood cell" dipolar BEC~\cite{Ronen07}.  Similarly, anisotropic lattice confinement has surprising differences from preconceptions based on uniform isotropic systems.  Dipolar interactions in confined geometries appear in many other branches of physics, such as in ferromagnetic nanostructures~\cite{EO_Demond_01}, where the dipole-dipole interaction plays a key role in the dispersion relation for spin waves in ferromagnetic films~\cite{Kalinikos_Slavin_86} and wires~\cite{Park_Eames_02}.

An important application of our results is in deriving many-body models to describe the low-energy physics of dipolar gases on lattices.  Previous derivations, for example Refs.~\cite{Pollet_Picon_10,capogrosso2010,Wall_Carr_10,Gorshkov_Manmana_11,Gorshkov_Manmana_11b}, assume that the interaction between localized lattice states has the same functional form as in the continuum.  Since this amounts to replacing the localized single-particle probability distributions with delta functions, we will call this the delta function approximation (DFA).  Performing matrix product state simulations~\cite{Schollwoeck_11} on infinite one-dimensional (1D) lattices, we demonstrate that the DFA can lead to significant errors in the determination of the phase diagram.

This paper is organized as follows.  In Sec.~\ref{sec:OL} we review the theory of Wannier functions and their use in deriving effective many-body lattice models for strongly correlated systems.  In particular, we provide a quantitative analysis of the decay properties of Wannier functions as well as properties of their squares interpreted as probability distributions.  Sec.~\ref{sec:Edd} provides numerical results for the effective dipole-dipole interactions in the presence of an optical lattice and discusses the effects of confinement.  In Sec.~\ref{sec:MBP} we study the phase diagram of hard-core bosons in one dimension with infinite-size matrix product state techniques to exemplify the impact of the confinement-induced modification of dipole-dipole interactions on many-body physics.  Finally, in Sec.~\ref{sec:Concl}, we conclude.   In \ref{sec:Numerical} we discuss numerical methods for evaluating matrix elements of nonlocal potentials in a basis of Wannier functions, and in \ref{sec:DDIEnergy}, we provide an explicit evaluation of the dipole-dipole interaction in a cylindrically symmetric harmonic trap for a comparison with the results in an optical lattice.

\section{Optical Lattices, Wannier functions, and Hubbard models}
\label{sec:OL}

The theory of quantum mechanical objects in a continuous periodic potential is well established~\cite{bloch2008}.  Here, we present a review of the basic facts in order to set notation, and also provide some explicit computations for the optical lattice potential Eq.~(\ref{eq:OLPotential}) which do not appear elsewhere in the literature, to the best of our knowledge.

We consider that our system is subject to the separable simple cubic potential
\begin{eqnarray}
\label{eq:OLPotential}V(\mathbf{r})&=\textstyle{\sum_{\nu\in\{x,y,z\}}V_{\nu}\sin^2(\pi \nu/a)}\, ,
\end{eqnarray}
where $a$ is the lattice spacing.  The typical energy scale derived from $a$ is the recoil energy $E_R=\hbar^2\pi^2/2ma^2$.  A potential of the form Eq.~\eref{eq:OLPotential} is produced for ultracold gases by an optical lattice consisting of three sets of counter-propagating laser beams.  The interaction of an optical potential which is far detuned from any atomic or molecular resonances may be described by the AC Stark shift~\cite{Bonin_Kresin_97}
\begin{eqnarray}
\hat{H}_{\mathrm{opt}}\left(\mathbf{r}\right)&=&-\mathbf{E}_{\mathrm{opt}}^{\star}\left(\mathbf{r}\right)\cdot \tilde{\alpha}\left(\omega_{\mathrm{opt}}\right)\cdot \mathbf{E}_{\mathrm{opt}}\left(\mathbf{r}\right)\, ,
\end{eqnarray}
where $\mathbf{E}_{\mathrm{opt}}\left(\mathbf{r}\right)$ is the optical field and $\tilde{\alpha}\left(\omega_{\mathrm{opt}}\right)$ is the dynamical polarizability tensor of the object evaluated at the optical frequency $\omega_{\mathrm{opt}}$.  The quantities $V_x$, $V_y$, and $V_z$ in Eq.~\eref{eq:OLPotential}, which we will refer to as the \emph{lattice heights} along the $x$, $y$, and $z$ directions, may be tuned by increasing the intensity of the optical field.  Throughout this paper, we will use the notation $\tilde{V}\equiv V/E_R$ to denote the dimensionless ratio of a lattice height to the recoil energy.  The assumption of a separable potential such as Eq.~\eref{eq:OLPotential} applies when the polarizability tensor is a scalar, as occurs naturally for alkali atoms.  When the dynamical polarizability contains non-scalar components, internal states of, e.g., a rotating molecule can be coupled together for certain polarizations of the optical potential~\cite{Neyenhuis_Yan_12}.  The theory given in this paper can be extended to this case, but the analysis is more complex.  For clarity of exposition, we will focus on the separable case, Eq.~\eref{eq:OLPotential}.

For particles subject to a periodic lattice potential, the energy eigenfunctions are Bloch functions $\psi_{\mathbf{nk}}(\mathbf{r})$ characterized by a quasimomentum index $\mathbf{k}$ in the first Brillouin zone (BZ) and a band index $\mathbf{n}$.  Bloch functions are the simultaneous eigenfunctions of the single-particle Hamiltonian and the lattice translation operators, and so represent the analogs of plane waves in free space when the translational symmetry is a discrete, rather than continuous, group.  As such, Bloch functions are delocalized objects, and so are often not an appropriate basis for expanding a many-body Hamiltonian with strong local interactions.  A more appropriate basis for describing strong interactions in lattices is provided by Wannier functions, which are the quasimomentum Fourier transforms of the Bloch functions,
\begin{eqnarray}
\label{eq:wanndef}w_{\mathbf{in}}(\mathbf{r})\equiv w_{\mathbf{n}}(\mathbf{r}-\mathbf{r}_{\mathbf{i}})&=\frac{1}{\sqrt{L}}\sum_{\mathbf{k}\in\mathrm{BZ}}e^{-i\mathbf{k}\cdot\mathbf{r}_i}\psi_{\mathbf{nk}}(\mathbf{r})\, .
\end{eqnarray}
Here $L$ is the total number of unit cells in a lattice with periodic boundary conditions and $\mathbf{r}_{\mathbf{i}}$ denotes the coordinate of lattice site $\mathbf{i}$.  For simplicity, we will restrict our attention to the lowest band in most cases, and drop the band index $\mathbf{n}$.  The extension of our results to multi-band situations~\cite{Dutta_Sowi_13} is straightforward.  Additionally, our methods also extend readily to other bases, for example localized bases which take into account strong contact interactions~\cite{Li_Yu_06,Hazzard_Mueller_10,Dutta_Eckardt_11,Bissbort_Deuretzbacher_12}.

\subsection{Properties of Wannier functions}
It was shown in a seminal work by Kohn~\cite{Kohn_59} that the phases on the Bloch functions can be chosen such that the Wannier functions are exponentially decaying away from their centers for one-dimensional centro-symmetric lattice potentials in a sense to be discussed in the next paragraph.  The Wannier functions with this choice of phases are called maximally localized Wannier functions, and are used throughout this paper.  The exponential decay of Wannier functions has been extended to general one-dimensional lattice potentials~\cite{Cloizeaux_64}, non-degenerate bands in arbitrary dimensions~\cite{Nenciu}, and general two- and three-dimensional insulators with vanishing Chern number~\cite{Brouder_Panati_07}.  Hence, the exponential localization of Wannier functions, which will play an important role in our results, is a general property which does not require fine tuning or a specific lattice structure.  

\begin{figure}[tbp]
\centerline{\includegraphics[width=1.0\columnwidth]{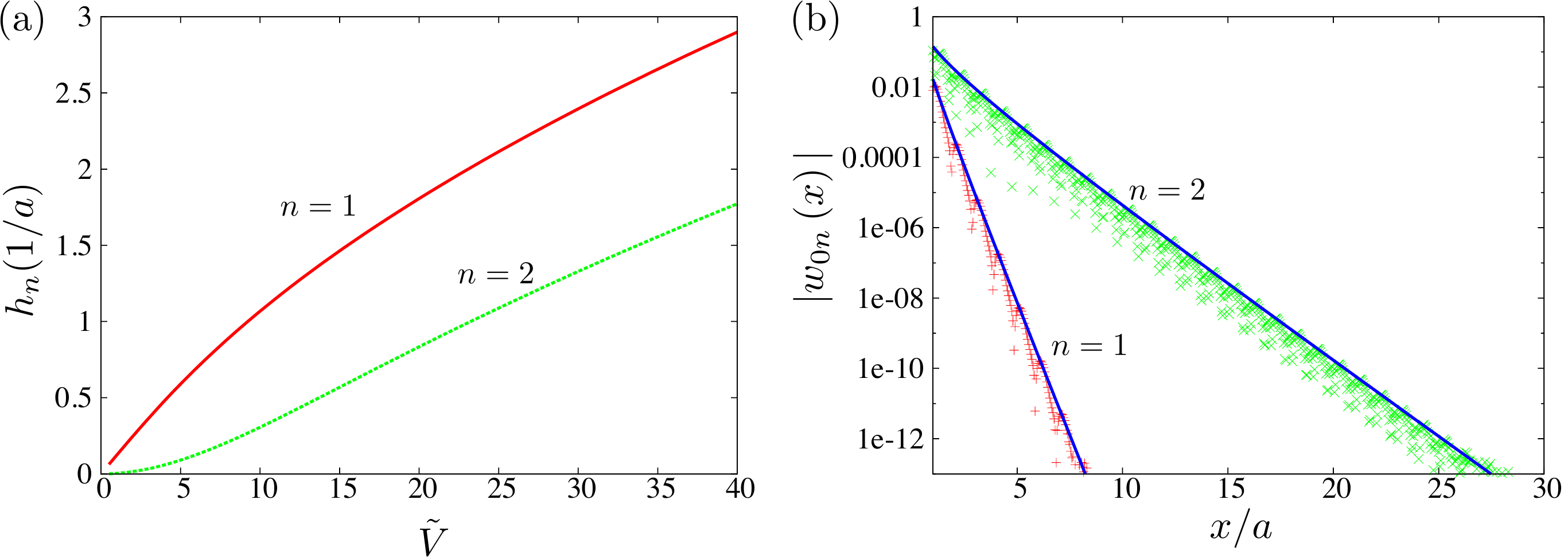}}
\caption{\label{fig:Wannierdecay} Panel (a): exponential decay constant $h_n$ of the 1D Wannier function as a function of the lattice height $\tilde{V}$ for the lowest two bands.  Deeper lattices localize the Wannier functions more effectively.  Panel (b): Decay of the Wannier functions of the first two bands (points) together with their fits to the asymptotic exponential form Eq.~\eref{eq:ActualWannierAsyp}(solid lines), $V=10E_R$.}
\end{figure}

Because the potential we consider, Eq.~\eref{eq:OLPotential}, is separable, the Bloch solutions of the 3D single-particle Schr\"odinger equation are products of the Bloch solutions of the 1D Hamiltonian
\begin{eqnarray}
\label{eq:1DHami}{H}\left(x\right)&=&-\frac{\hbar^2}{2m}\frac{d^2}{dx^2}+V\sin^2\left(\frac{\pi x}{a}\right)\, .
\end{eqnarray}
The linearity of the transformation Eq.~\eref{eq:wanndef} implies that the 3D Wannier functions of Eq.~\eref{eq:OLPotential} are hence products of 1D Wannier functions obtained by quasimomentum Fourier transform of the 1D Bloch eigenfunctions of Eq.~\eref{eq:1DHami},
\begin{eqnarray}
w_{\mathbf{in}}\left(\mathbf{r}\right)&=&w_{i_xn_x}\left(x\right)w_{i_yn_y}\left(y\right)w_{i_zn_z}\left(z\right)\, .
\end{eqnarray}
Thus, we can use the results of Kohn~\cite{Kohn_59} to discuss the properties of the 1D Wannier functions $w_{in}\left(x\right)$.  In particular, the Wannier functions decay exponentially as $|w_{0n}\left(x\right)|\sim \exp(-h_nx)$ in the sense that
\begin{eqnarray}
\label{eq:Wannierdecay}\lim_{x\to\infty} |w_{0n}\left(x\right)|e^{qx}&=&\left\{\begin{array}{c} 0\, ,\;\;q<h_n\\\infty\, ,\;\;q>h_n\end{array}\right.\, .
\end{eqnarray}
The parameter $h_n$ is the distance to the nearest branch point from the real axis in the complex quasimomentum plane.  The decay length $h_n$ can be determined numerically by locating crossings in the band structure computed with quasimomentum $k=k_n+i\kappa$, $k_n=\pi (1-\left(-1\right)^{n+1} )/2a$ and $\kappa$ real.  At the point $k$ where bands $n$ and $n+1$ cross, $h_n=\kappa$.  The results of this analysis for the lowest two bands of the 1D potential Eq.~\eref{eq:1DHami} are shown in Fig.~\ref{fig:Wannierdecay}(a).  He and Vanderbilt~\cite{He_Vanderbilt_01} pointed out that Eq.~\eref{eq:Wannierdecay} is consistent with an exponential decay multiplied by an algebraic factor.  They then demonstrated that the precise asymptotic behavior of maximally localized, orthogonal Wannier functions in 1D is
\begin{eqnarray}
\label{eq:ActualWannierAsyp} |w_{0n}\left(x\right)|&\sim&\exp\left(-h_nx\right)x^{-3/4}\, ,
\end{eqnarray}
for a lattice consisting of periodically repeating Gaussian wells.  By performing numerical fits with the obtained $h_n$ from Fig.~\ref{fig:Wannierdecay}(a), we find that this same algebraic factor appears in the Wannier functions of the potential Eq.~\eref{eq:1DHami}, and the power of the algebraic correction does not appear to depend on the lattice height or the band index.  The resulting fits to the Wannier functions, together with the numerically computed Wannier functions, are shown in Fig.~\ref{fig:Wannierdecay}(b).  Fig.~\ref{fig:Wannierdecay} provides a rigorous definition of the decay of the Wannier functions we consider in this paper as a function of the height of the optical lattice potential.

\subsection{Approximation of Wannier functions by Harmonic oscillator functions}
\label{sec:hoapprox}
It is difficult to obtain quantitative analytical predictions from Wannier functions due to their complicated form.  Hence, approximations to the true Wannier functions are often made for analytical convenience.  The most common such approximation is to replace a single site of the optical lattice by a harmonic well with the same local curvature.  We shall call this the harmonic oscillator approximation (HOA).  The curvature-matching condition amounts to $\ell_{\nu}=a/(\pi \tilde{V}_{\nu}^{1/4})$, where $\ell_{\nu}$ is the harmonic oscillator length along Cartesian direction $\nu$.  The 1D ground state wave function, which is taken to approximate the 1D lowest band Wannier function, is
\begin{eqnarray}
\label{eq:hoapprox} \psi_{\mathrm{ho}}\left(x\right)&=\frac{1}{\sqrt{\ell\sqrt{\pi}}}\exp\left(-\frac{x^2}{2\ell^2}\right)\, .
\end{eqnarray}
There are several important differences between the HOA Eq.~(\ref{eq:hoapprox}) and the true Wannier function.  First, Eq.~(\ref{eq:hoapprox}) is everywhere positive, while Wannier functions, even for the lowest band, always have nodes in order to maintain orthogonality between lattice sites.  Thus, the HOA poorly captures quantities like tunneling, which involve overlaps of derivatives of Wannier functions.  Also, Eq.~(\ref{eq:hoapprox}) decays much more rapidly than the true Wannier functions, as a Gaussian rather than an exponential.  Hence, the HOA will consistently underestimate the overlap of Wannier functions at different sites.  Third, the HOA Eq.~(\ref{eq:hoapprox}) is more peaked around its center value than the true Wannier function, and so quantities computed on-site are overestimated by the HOA.  Finally, the true Wannier functions have all of the lattice symmetries, while the HOA displays the symmetries of an ellipsoid for the simple cubic lattice.  The difference in symmetries is especially important when discussing interaction-induced diagonal tunneling~\cite{Wall_Carr_12,Wall_Carr_13}. 

\begin{figure}[tbp]
\centerline{\includegraphics[width=0.5\columnwidth]{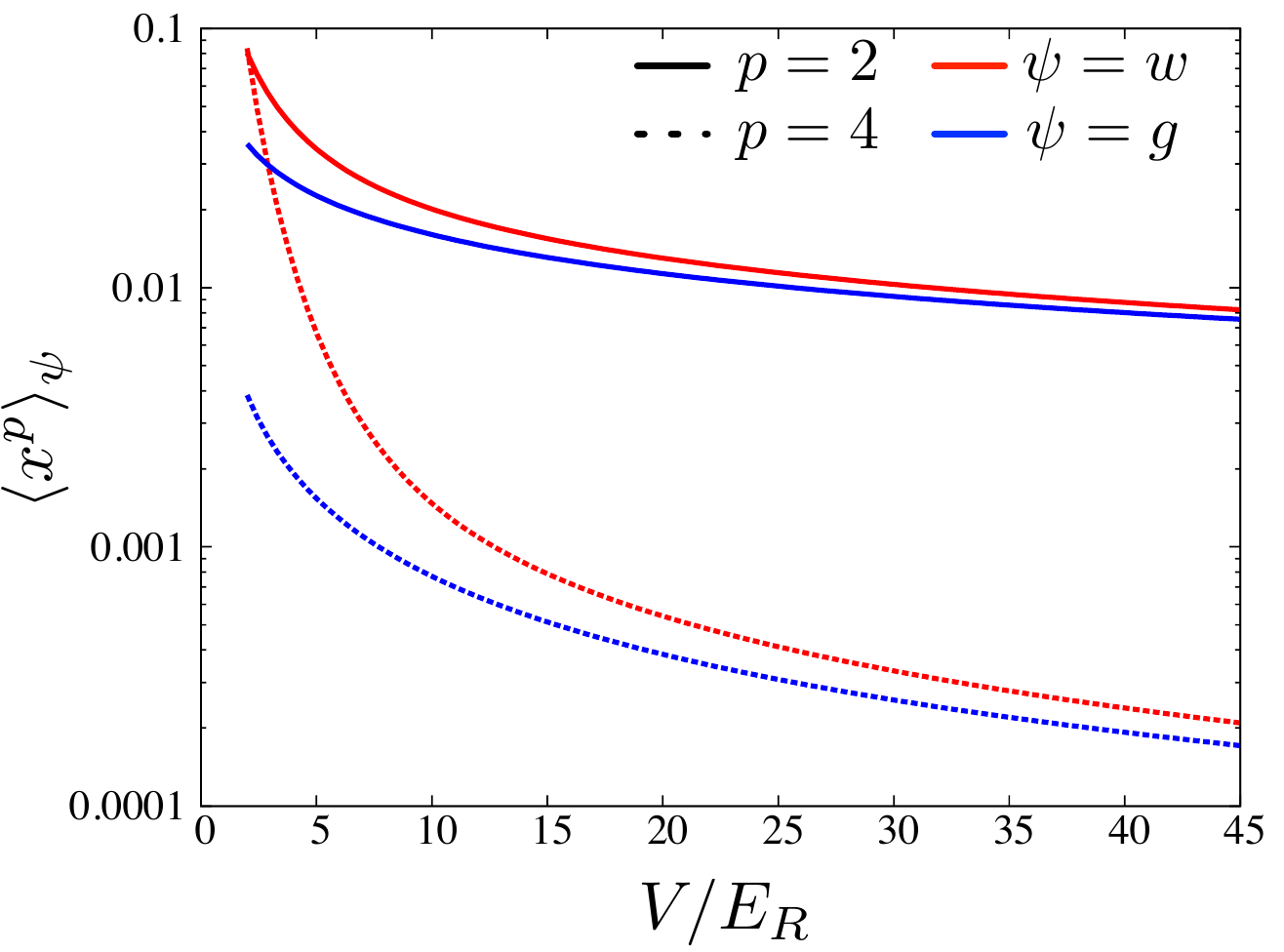}}
\caption{\label{fig:moments} The second (solid lines) and fourth (dashed lines) moments of the lowest band Wannier function (red) and the HOA, Eq.~(\ref{eq:hoapprox}), (blue) as a function of the lattice height.  While the HOA captures the mean width of the Wannier functions for deep lattices, the heavy tails of the Wannier functions cause large deviations in the fourth moments, especially in shallow lattices.}
\end{figure}

We can make the comparison between Wannier functions and the HOA more quantitative by considering the moments  of their associated single-particle probability distributions.  We define the $p^{th}$ one-dimensional moments as
\begin{eqnarray}
\langle x^p\rangle_{\psi}&\equiv \left\{\begin{array}{c} \int_{-\infty}^{\infty} dx \left|w\left(x\right)\right|^2 x^p\,, \;\;\; \psi=w\\ \frac{1}{\sqrt{\pi}\ell}\int_{-\infty}^{\infty} dx \exp\left(-\frac{x^2}{\ell^2}\right)x^p\, ,\;\;\; \psi=g\end{array}\right.\, ,
\end{eqnarray}
where $\psi=w$ and $g$ stand for Wannier and Gaussian, respectively.  The second moment gives an estimate for the width of the distribution.  It is convenient to define the kurtosis $\kappa_\psi\equiv (\langle x^4\rangle_\psi/\langle x^2\rangle_\psi^2)-3$, which is a measure of the peakedness of a distribution as well as the heaviness of its tails.  We show the second and fourth moments of the true Wannier functions and the HOA in Fig.~\ref{fig:moments}.  The width of the lowest band Wannier function is well captured by the HOA for deep lattices with $\tilde{V}\gtrsim 20$, although the HOA always has a smaller second moment than the true Wannier function.  The harmonic oscillator always has a vanishing kurtosis, $\kappa_g=0$, as is known for Gaussian distributions.  In contrast, the Wannier functions have an always positive kurtosis which is very sizable for shallow lattices and approaches zero for deeper lattices.  For example, we observed $\kappa_w=10$ for $\tilde{V}=2$ and $\kappa_w=0.12$ for $\tilde{V}=35$.  This difference in the kurtosis quantifies the essential difference between the decay of the Wannier functions and the HOA.

\subsection{Derivation of Hubbard models for dipolar particles}
With the identification of the lowest band Wannier functions as the appropriate single-particle basis for describing strongly interacting particles in a lattice, we derive a many-body lattice model using the well-known procedure~\cite{jaksch1998,lewensteinM2007} of expanding the field operator $\hat{\psi}(\mathbf{r})$ in the basis of Wannier functions and substituting this expansion into the second-quantized expression for the interaction Hamiltonian,
\begin{eqnarray}
\label{eq:SQH} \hat{H}_{\mathrm{int}}&={\frac{1}{2}\int d\mathbf{r}\int d\mathbf{r}'\,}\hat{\psi}^{\dagger}(\mathbf{r})\hat{\psi}^{\dagger}(\mathbf{r}'){V}_{\mathrm{int}}(\mathbf{r}-\mathbf{r}')\hat{\psi}(\mathbf{r}')\hat{\psi}(\mathbf{r})\, ,
\end{eqnarray}
where ${V}_{\mathrm{int}}\left(\mathbf{r}\right)$ is the two-particle interaction potential.  The expansion of Eq.~\eref{eq:SQH} in Wannier functions yields
\begin{eqnarray}
\label{eq:HWannierExp}\hat{H}_{\mathrm{int}}&=\frac{1}{2}\sum_{\mathbf{i}_1\mathbf{i}_2\mathbf{i}_2'\mathbf{i}_1'}\mathcal{U}_{\mathbf{i}_1\mathbf{i}_2\mathbf{i}_2'\mathbf{i}_1'}\hat{a}_{\mathbf{i}_1}^{\dagger}\hat{a}_{\mathbf{i}_2}^{\dagger}\hat{a}_{\mathbf{i}_2'}\hat{a}_{\mathbf{i}_1'}\, ,
\end{eqnarray}
where $\hat{a}_{\mathbf{i}}$ destroys a particle in a lowest band Wannier state centered at site $\mathbf{i}$ and
\begin{eqnarray}
\mathcal{U}_{\mathbf{i}_1\mathbf{i}_2\mathbf{i}_2'\mathbf{i}_1'}&\equiv{\int d\mathbf{r}\int d\mathbf{r}'\,} f_{\mathbf{i}_1\mathbf{i}_1'}(\mathbf{r}){V}_{\mathrm{int}}(\mathbf{r}-\mathbf{r}')f_{\mathbf{i}_2\mathbf{i}_2'}(\mathbf{r})\, .
\end{eqnarray}
Here, $f_{\mathbf{i}\mathbf{i}'}(\mathbf{r})\equiv w_{\mathbf{i}}^{\star}(\mathbf{r})w_{\mathbf{i}'}(\mathbf{r})$ is a product of Wannier functions.  The matrix elements $\mathcal{U}_{\mathbf{i}_1\mathbf{i}_2\mathbf{i}_2'\mathbf{i}_1'}$, which we will call \emph{Hubbard parameters}, describe the interactions between particles localized in lowest band Wannier states.

The interaction we consider is the dipole-dipole interaction, commonly given as
\begin{eqnarray}
V_{\mathrm{DD}}\left(\mathbf{r}\right)&=\frac{1}{r^3}\left[\hat{\mathbf{d}}\cdot \hat{\mathbf{d}}-3\left(\hat{\mathbf{d}}\cdot\mathbf{e}_r\right)\left(\mathbf{e}_r\cdot\hat{\mathbf{d}}\right)\right]\, ,
\end{eqnarray}
where $\hat{\mathbf{d}}$ is the dipole operator, $\mathbf{r}$ is the relative position of the interacting particles, $r=\left|\mathbf{r}\right|$, and $\mathbf{e}_r$ is a unit vector in the direction of $\mathbf{r}$.  It is convenient to instead recast the dipole-dipole potential as the contraction of two spherical tensors as~\cite{Zare}
\begin{eqnarray}
V_{\mathrm{DD}}\left(\mathbf{r}\right)&=-\frac{\sqrt{6}}{r^3}\sum_{q=-2}^{2}\left(-1\right)^{q}C^{\left(2\right)}_{-q}\left(\mathbf{r}\right)\left[\hat{\mathbf{d}}\otimes \hat{\mathbf{d}}\right]^{\left(2\right)}_{q}\, ,
\end{eqnarray}
where $C^{\left(2\right)}_q\left(\mathbf{r}\right)=\sqrt{\frac{4\pi}{5}}Y^{\left(2\right)}_{q}\left(\mathbf{r}\right)$ is an unnormalized spherical harmonic in the spherical coordinates of $\mathbf{r}$ and
\begin{eqnarray}
\left[\hat{\mathbf{d}}\otimes \hat{\mathbf{d}}\right]^{\left(2\right)}_{q}&=&\sum_m\langle 1,m,1,q-m|2,q\rangle \hat{d}_{m}\otimes \hat{d}_{q-m}
\end{eqnarray}
is the irreducible tensor product of two dipole operators with $\langle j_1m_2j_2m_2|jm\rangle$ a Clebsch-Gordan coefficient.  The operators
\begin{eqnarray}
\hat{d}_{\pm 1}&\equiv &\mp \left(\hat{d}_x\pm i\hat{d}_y\right)/\sqrt{2}\, ,\;\;\hat{d}_0\equiv \hat{d}_z\, ,
\end{eqnarray}
are the spherical decompositions of the vector operator $\hat{\mathbf{d}}$.  The interaction Hamiltonian may be written in terms of coupling constants $U^{\mathrm{DD},q}$ and geometrical factors $\mathcal{G}^{\mathrm{DD},q}$ as
\begin{eqnarray}
\label{eq:DDIInt}\hat{H}_{\mathrm{int}}&=\frac{1}{2}\sum_{q=-2}^{2}\left(-1\right)^{q}U^{\mathrm{DD},-q}\sum_{\mathbf{i}_1\mathbf{i}_2\mathbf{i}_2'\mathbf{i}_1'}\mathcal{G}^{\mathrm{DD},q}_{\mathbf{i}_1\mathbf{i}_2\mathbf{i}_2'\mathbf{i}_1'}\hat{a}_{\mathbf{i}_1}^{\dagger}\hat{a}_{\mathbf{i}_2}^{\dagger}\hat{a}_{\mathbf{i}_2'}\hat{a}_{\mathbf{i}_1'}\, ,
\end{eqnarray}
where
\begin{eqnarray}
\label{eq:DDCC}U^{\mathrm{DD},q}&\equiv& \sqrt{\frac{3}{2}}{\left[\hat{\mathbf{d}}\otimes \hat{\mathbf{d}}\right]^{\left(2\right)}_{q}}/{a^3}\, ,\\
\label{eq:DDGeo}\mathcal{G}^{\mathrm{DD},q}_{\mathbf{i}_1\mathbf{i}_2\mathbf{i}_2'\mathbf{i}_1'}&\equiv &-2a^3{\int d\mathbf{r}\int d\mathbf{r}'\,} f_{\mathbf{i}_1\mathbf{i}_1'}(\mathbf{r})\frac{C^{\left(2\right)}_{q}(\mathbf{r}-\mathbf{r}')}{|\mathbf{r}-\mathbf{r}'|^3}f_{\mathbf{i}_2\mathbf{i}_2'}(\mathbf{r}')\, .
\end{eqnarray}
All of the information about the size of the dipole moment and the lattice spacing are contained in the coupling constant Eq.~(\ref{eq:DDCC}).  On the other hand, the geometrical integral Eq.~(\ref{eq:DDGeo}) contains the information about the effects of lattice confinement on the effective interaction through the Wannier function products $f_{\mathbf{ii}'}\left(\mathbf{r}\right)$.  It is these geometrical factors which we are interested in studying in the present article.

While our methods can be applied to all components $q$ of the dipole-dipole interaction, for simplicity we focus on the terms in Eq.~(\ref{eq:DDIInt}) with $q=0$.  These terms do not change the projection of the total angular momentum along a space-fixed quantization axis.  The $q=0$ components are the only terms relevant for ultracold $^1\Sigma$ molecules in an optical lattice and oriented in a strong DC electric field, as states with dipole-allowed transitions are separated by energy splittings large compared to the characteristic dipole-dipole interaction energy~\cite{MHHPRL,Gorshkov_Manmana_11}.  Also, these are the most relevant processes for magnetic dipoles in an optical lattice with a strong magnetic field to prevent dipolar relaxation and spontaneous demagnetization~\cite{Pasquiou_Bismut_10,Pasquiou_Marechal_106,Pasquiou_Marechal_108}, or a quantum simulation of magnetic dipoles using symmetric top molecules in a strong electric field~\cite{Wall_Maeda_13}.  In this case, we have
\begin{eqnarray}
U^{\mathrm{DD},0}&=&\frac{1}{a^3}\left[\hat{d}_0\hat{d}_0+\frac{1}{2}\left(\hat{d}_1\hat{d}_{-1}+\hat{d}_{-1}\hat{d}_1\right)\right]\, ,\\
\mathcal{G}^{\mathrm{DD},0}_{\mathbf{i}_1\mathbf{i}_2\mathbf{i}_2'\mathbf{i}_1'}&=&-2a^3{\int d\mathbf{r}\int d\mathbf{r}'\,} f_{\mathbf{i}_1\mathbf{i}_1'}(\mathbf{r})\frac{C^{\left(2\right)}_{0}(\mathbf{r}-\mathbf{r}')}{|\mathbf{r}-\mathbf{r}'|^3}f_{\mathbf{i}_2\mathbf{i}_2'}(\mathbf{r}')\\
&=&a^3{\int d\mathbf{r}\int d\mathbf{r}'\,} f_{\mathbf{i}_1\mathbf{i}_1'}(\mathbf{r})\frac{1-3\cos^2\theta}{|\mathbf{r}-\mathbf{r}'|^3}f_{\mathbf{i}_2\mathbf{i}_2'}(\mathbf{r}')\, ,
\end{eqnarray}
where $\theta$ is the polar angle of the relative coordinate $(\mathbf{r}-\mathbf{r}')$.  The commonly used delta-function approximation (DFA)~\cite{Pollet_Picon_10,capogrosso2010,Wall_Carr_10,Gorshkov_Manmana_11,Gorshkov_Manmana_11b} replaces $f_{\mathbf{ii}'}\left(\mathbf{r}\right)\to \delta\left(\mathbf{r}-\mathbf{r}_{i}\right)\delta_{\mathbf{ii}'}$, such that
\begin{eqnarray}
\mathcal{G}^{\mathrm{DD},0}_{\mathbf{i}_1\mathbf{i}_2\mathbf{i}_2'\mathbf{i}_1'}&=&\delta_{\mathbf{i}_1\mathbf{i}_1'}\delta_{\mathbf{i}_2\mathbf{i}_2'}a^3\frac{1-3\cos^2\theta_{i_1i_2}}{\left|\mathbf{r}_{i_1}-\mathbf{r}_{i_2}\right|^3}\, ,
\end{eqnarray}
where $\theta_{i_1i_2}$ is the polar angle of the relative vector between lattice sites $\mathbf{i}_1$ and $\mathbf{i}_2$ with positions $\mathbf{r}_{i_1}$ and $\mathbf{r}_{i_2}$, respectively.

The terms in Eq.~(\ref{eq:DDIInt}) may be viewed as scattering processes in which particles in the lowest band at lattice sites $\mathbf{i}_1'$ and $\mathbf{i}_2'$ move to lattice sites $\mathbf{i}_1$ and $\mathbf{i}_2$, respectively, in the course of an interaction.  The largest magnitude terms in Eq.~(\ref{eq:DDIInt}) are those in which $\mathbf{i}_1=\mathbf{i}_1'$ and $\mathbf{i}_2=\mathbf{i}_2'$, which we will call \emph{direct interactions} using the scattering process analogy.  The direct terms arise as density-density interactions in the interaction Hamiltonian, proportional to $\hat{n}_{i_1}\hat{n}_{i_2}$.  Another class of terms which are proportional to $\hat{n}_{i_1}\hat{n}_{i_2}$ are the \emph{exchange interactions}, in which $\mathbf{i}_1=\mathbf{i}_2'$ and $\mathbf{i}_2=\mathbf{i}_1'$ with $\mathbf{i}_1'\ne \mathbf{i}_2'$.  All other processes involve changing the position of a Wannier state from its initial position.  Examples of such processes are assisted tunneling~\cite{Gorshkov_Manmana_11b} and pair hopping~\cite{Wall_Carr_12,Sowi_Dutta_12}, both of which can introduce new quantum phases in shallow lattices~\cite{Sowi_Dutta_12}.  However, assisted tunneling, pair hopping, and exchange terms are all suppressed by factors related to the exponential decay of the Wannier functions, see Fig.~\ref{fig:Wannierdecay}, and the behavior of these Hubbard parameters with lattice confinement is qualitatively similar.  Because the focus of this work is on the qualitative behavior of the Hubbard parameters with the lattice confinement, we will restrict our attention to the direct and exchange terms with the understanding that the behavior of assisted tunneling and pair hopping is similar to that of the exchange term.  It should be noted that our methods apply to any Hubbard parameter.  Hence, keeping only the direct and exchange interactions and taking into account the statistics of the particles, the expansion Eq.~\eref{eq:HWannierExp} may be written for the dipole-dipole interaction as
\begin{eqnarray}
\label{eq:Effint}  \hat{H}_{\mathrm{int}}&=U[\frac{1}{2} \sum_{\mathbf{i}} I_{0}\hat{n}_{\mathbf{i}}(\hat{n}_{\mathbf{i}}-1)+\frac{1}{2}\sum_{\mathbf{i}\ne\mathbf{i}'}I_{\mathbf{i}',\mathbf{i}}\hat{n}_{\mathbf{i}}\hat{n}_{\mathbf{i}'}]\, .
\end{eqnarray}
where we have defined $U\equiv U^{\mathrm{DD},0}$ and
\begin{eqnarray}
\label{eq:UHubb} I_{0}&\equiv \textstyle \mathcal{G}^{\mathrm{DD};0}_{\mathbf{0000}}\, ,\;\;\textstyle I_{\mathbf{i}',\mathbf{i}}\equiv \textstyle [\mathcal{G}^{\mathrm{DD};0}_{\mathbf{ii'i'i}}\pm \mathcal{G}^{\mathrm{DD};0}_{\mathbf{ii'ii'}}]\, .
\end{eqnarray}
The plus (minus) sign on the exchange term in Eq.~\eref{eq:UHubb} refers to bosons (fermions).  Results for the effective interactions, Eq.~(\ref{eq:UHubb}), computed using the methods of \ref{sec:Numerical}, are presented in the next section.

\section{Effective dipole-dipole interactions}
\label{sec:Edd}

With respect to the length scale of the lattice constant $a$, the short range physics is given by the interactions $I_0$ which occur within a unit cell.  A point of comparison for the on-site dipolar interaction $I_0$ is provided by the dimensionless interaction energy $I_0^{\mathrm{ho}}=\langle \hat{H}_{\mathrm{DD}}\rangle /U$ of two bosonic particles in the ground state of a harmonic trap, where the harmonic trap is chosen to match the local curvature of a lattice site as detailed in Sec.~\ref{sec:hoapprox}.  In the case where the oscillator lengths in the $xy$ plane are equal, $\ell_x=\ell_y=\ell_{\perp}$, we have that
\begin{eqnarray}
\label{eq:U0ho}I_0^{\mathrm{ho}}&= \frac{\sqrt{2}}{\pi  \bar{\ell}\,^3}\left[\frac{2}{3}+\beta^2-\frac{\beta}{1-\alpha^2}\cot^{-1}(\beta)\right] \, ,
\end{eqnarray}
where $\bar{\ell}\equiv (\ell_z\ell_{\perp}^2)^{1/3}/a$ is the geometric mean oscillator length in units of $a$, $\alpha\equiv \ell_z/\ell_{\perp}$ measures the confinement asymmetry, and $\beta\equiv \alpha (1-\alpha^2)^{-1/2}$.  Notably, the interaction energy $I_0^{\mathrm{ho}}$ vanishes for isotropic confinement, $\alpha=1$.  For $\alpha<1$, corresponding to stronger confinement along the quantization axis, contributions from $\theta> \arccos \sqrt{\frac{1}{3}}$, where the dipole-dipole potential is negative, are suppressed.  Hence, $I_0^{\mathrm{ho}}$ is positive for $\alpha<1$.  In contrast, for $\alpha>1$ where confinement is weakest along the quantization axis, $I_0^{\mathrm{ho}}$ is negative.  These qualitative features are shared by the true effective interaction using Wannier functions, as is shown in \ref{sec:DDIEnergy}.  Of particular interest is that the on-site dipole-dipole interaction vanishes when the lattice heights are equal, $\tilde{V}\equiv \tilde{V}_x=\tilde{V}_y=\tilde{V}_z$.  Also, as noted in Sec.~\ref{sec:hoapprox}, the HOA energy $I_0^{\mathrm{ho}}$ is always greater than the on-site energy computed using Wannier functions.

For the moderate- to long-range interactions $I_{\mathbf{i}',\mathbf{i}}$, let us define $I_j\equiv I_{\mathbf{i}+\mathbf{j},\mathbf{i}}$ with $\mathbf{j}$ along the $x$ direction.  The parameters $I_j$ correspond to effective dipole-dipole interactions separated by a distance of $j$ lattice spacings along a principal axis.  The DFA predicts that the exchange contribution vanishes and the direct term contributes a factor of $j^{-3}$ such that $I_j=j^{-3}$.  Hence, there are two possible sources of deviation from the DFA.  The first source is a non-vanishing contribution from the exchange term.  The second source is a deviation of the direct term from $j^{-3}$.  The exchange term is non-vanishing due to overlap between Wannier functions on different sites, and so decreases with increasing lattice height as the exponential localization factor $h_1$ increases, see Fig.~\ref{fig:Wannierdecay}.  We can expect that the exchange processes will be negligible in magnitude even at the nearest-neighbor distance when $2h_1 a>1$, as then the Wannier functions are well-localized within a unit cell.  From Fig.~\ref{fig:Wannierdecay}, this gives an estimation of $\tilde{V}\approx 5$.  In contrast to the modification of the effective interaction by the exchange term, modification of the effective interaction due to the direct term relies essentially on asymmetry in three-dimensional confinement.  This can be understood in analogy with the effective short-range interaction discussed in the last paragraph, and is a consequence of the fact that the dipole-dipole interaction has relative $d$-wave symmetry.

\begin{figure}[tbp]
\centerline{\includegraphics[width=1.0\columnwidth]{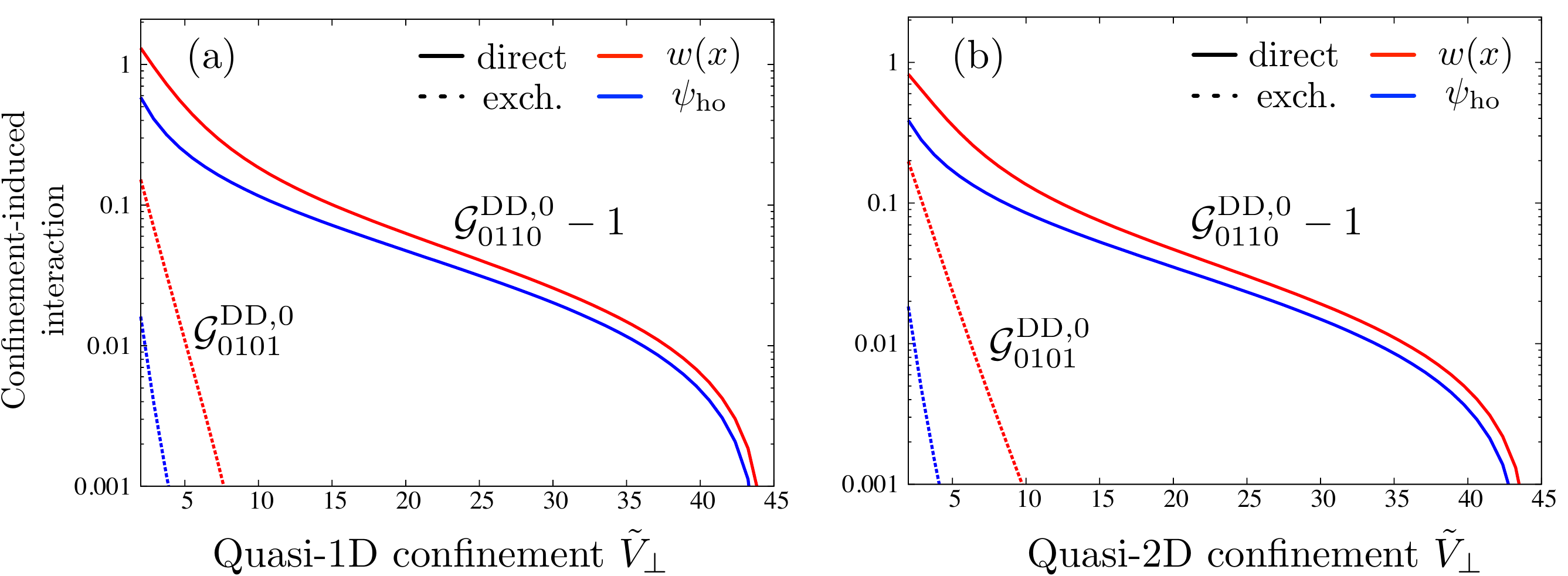}}
\caption{\label{fig:DvsE}  Panels (a) and (b) show the modifications of effective dipole-dipole interactions from their DFA values at the nearest-neighbor distance in quasi-1D and quasi-2D, respectively, as a function of confinement asymmetry.  The shown values are the dimensionless integrals Eq.~(\ref{eq:DDGeo}), but may also be interpreted as the confinement-induced interaction in units of $U^{\mathrm{DD},0}$ from Eq.~(\ref{eq:DDIInt}).  The solid lines are the deviations from the DFA prediction for the direct term, and the dashed lines are the exchange term.  Note that the exchange term identically vanishes in the DFA.  The red curves are the values computed with actual Wannier functions, and the blue curves use the HOA.}
\end{figure}

To demonstrate the effects of asymmetric lattice confinement, we evaluate the parameters {$I_{\mathbf{j}}$} for two quasi-low dimensional scenarios.  In the quasi-2D scenario, we take the $z$ direction to be tightly confined with a lattice height $\tilde{V}_z=45 $ and the lattice heights along the $x$ and $y$ directions to be equal, $\tilde{V}_{\perp}\equiv \tilde{V}_x=\tilde{V}_y$.  In the quasi-1D scenario {we take }both the $z$ and $y$ lattices {to be} tightly confining with $\tilde{V}_z=\tilde{V}_y=45$, and call the $x$ lattice height $\tilde{V}_{\perp}$.  These quasi-dimensional reductions are quite moderate and are commonplace in current experiments~\cite{Greiner_Mandel_02,Stoeferle_Moritz_04}.  The confinement-induced modifications of effective dipole-dipole interactions with respect to the DFA in the quasi-1D and quasi-2D scenarios are shown in Fig.~\ref{fig:DvsE}.  Here, the modifications to the direct term are shown as $(\mathcal{G}_{0110}-1)$, which is the deviation from the DFA prediction of 1 at the nearest-neighbor distance.  The exchange term is identically zero in the DFA, and so any non-vanishing value for the exchange term represents a deviation from the DFA.  The red curves are the numerical computation with Wannier functions, and the blue curves use the HOA.  We note that the exchange contributions, shown with dashed lines, are drastically underestimated by the HOA, as expected.  The modifications of the direct term, shown with solid lines, display qualitatively similar features between the true solution and the HOA.  In particular, both show vanishing modifications of the direct term when there is no asymmetry in confinement, $\tilde{V}_{\perp}=45$.  However, the HOA underestimates the modifications of the direct term, quite significantly for shallow quasi-low-dimensional confinement.

\begin{figure}[tbp]
\centerline{\includegraphics[width=0.8\columnwidth]{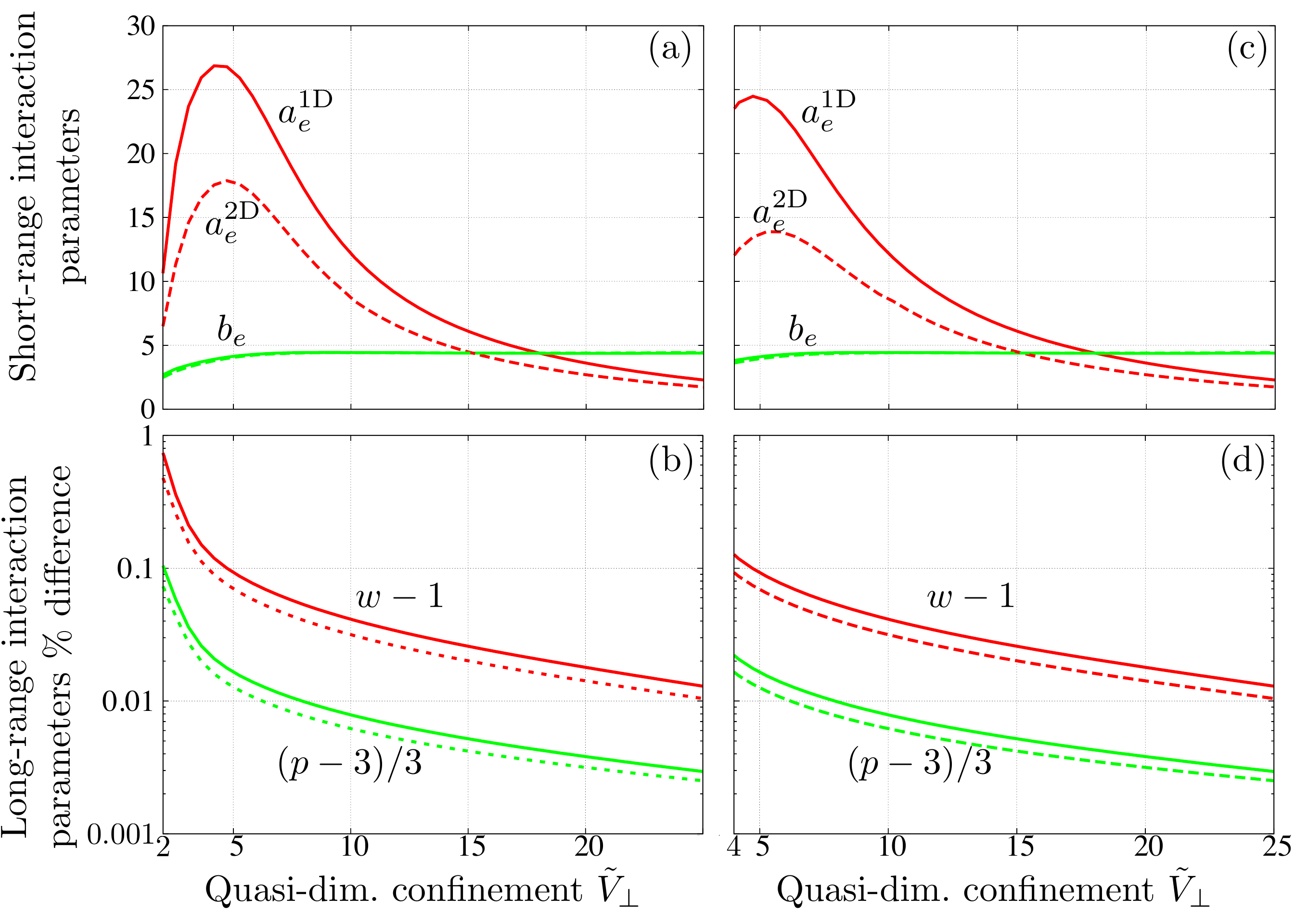}}
\caption{\label{fig:1D2DCompLog}  In all panels the solid (dashed) lines denote the best fit parameters to Eq.~\eref{eq:Uellfit} in the quasi-1D (quasi-2D) scenarios as a function of the quasi-low dimensional lattice height $\tilde{V}_{\perp}$.  Panels (a)-(b) pertain to bosons and panels (c)-(d) to fermions.  The top panels are the exponential weight $a_e$ (red) and the decay constant $b_e$ (green).  The bottom panels are percent differences of long-range weight $w$ (red) and power $p$ (green) with respect to the DFA.  Confinement effects are strongest for large confinement asymmetry, shallow quasi-dimensional confinement, and small separation between lattice sites.}
\end{figure}

In Fig.~\ref{fig:DvsE}, we showed the behavior of the confinement-induced modifications of the effective dipole-dipole interactions with confinement asymmetry at the nearest-neighbor distance.  These modifications also depend on the distance between sites in the lattice, and this dependence has a length scale set by the confinement.  In order to get a complete picture of the dependence of confinement effects on both distance between sites and confinement asymmetry we fit the numerically obtained data for {$I_{j}$} to the form
\begin{eqnarray}
\label{eq:Uellfit}I_{j}&=a_e\exp(-b_ej)+wj^{-p}\, ,
\end{eqnarray}
for $j\in[1,7]$, i.e., out to seven sites.  The DFA predicts no exponential, $a_e=0$ or $b_e\to\infty$, and $w=1$ and $p=3$ for the long-range contribution.  The fit parameters for our numerically generated data are shown in Fig.~\ref{fig:1D2DCompLog}.  Here the solid (dashed) lines refer to the quasi-1D (quasi-2D) scenarios.  The top panels are the short-range parameters $a_e$ and $b_e$ in Eq.~\eref{eq:Uellfit}, with {panel (a) (panel (c)) pertaining to bosons (fermions)}.  The bottom panels are the percent differences of the long-range parameters $w$ and $p$ in Eq.~\eref{eq:Uellfit} with respect to the DFA predictions.  Here, {panel (b) (panel (d)) pertains to bosons (fermions)}.  We note that for quasi-low dimensional confinement $\tilde{V}_{\perp}\gtrsim 7$, the predictions for bosons and fermions are the same to a few percent.  This implies that the exchange contribution in Eq.~\eref{eq:UHubb} plays no role for deep lattices, where deep lattices corresponds to $\tilde{V}\simeq 5$ in accordance with our expectations.

We chose the ansatz Eq.~\eref{eq:Uellfit} from several fit functions because it had the lowest fitting error and it provides a characteristic length scale $a_c\sim a/b_e$ of the confinement-induced modifications.  However, we do not propose that the form Eq.~\eref{eq:Uellfit} is exact.  We also stress that the exponential constant $b_e$ has no \emph{a priori} relation to the exponential decay constant $h_n$ of the Wannier functions.  Across a wide range of quasi-low dimensional confinement, $a_c\sim0.2 a$, and so the moderate range over which confinement modifies interactions is a few lattice sites.  It should also be noted that the confinement-induced modifications of the effective interactions will also have a nontrivial angular dependence due to the fact that Wannier functions are not spherically symmetric, but rather have the symmetries of the lattice.  We leave investigations of the nontrivial angular dependence for future work.

\section{Many-body physics}
\label{sec:MBP}

To illustrate the implications of our findings for many-body physics, we study a model of quasi-1D hard-core bosons with long-range dipolar interactions~\cite{MHHPRL}
\begin{eqnarray}
\label{eq:HCLRBH}\hat{H}&=\textstyle -t\sum_{\langle i j\rangle }\hat{a}_i^{\dagger}\hat{a}_{j}+U\sum_{i<j}I_{j-i}\hat{n}_i\hat{n}_j-\mu\sum_{i}\hat{n}_i\, .
\end{eqnarray}
Here, the nearest-neighbor tunneling amplitude is $t$, $\langle i,j\rangle$ denotes nearest-neighbor pairs $i$ and $j$, and $\mu$ is the chemical potential.  We compute the phase diagram of Eq.~\eref{eq:HCLRBH} for two realizations of {$I_{j}$} using the infinite size variational ground state search algorithm for matrix product states (iMPS)~\cite{McCulloch_08}.  In one realization, we use the DFA result, {$I_{j}=j^{-3}$}.  In the second realization, {$I_{j}$} is the effective interaction in an optical lattice computed with Wannier functions, Eq.~\eref{eq:UHubb}.  We take the lattice heights to be $\tilde{V}_{z}=\tilde{V}_{y}=45$, $\tilde{V}_{x}=6$, which fixes the tunneling energy $t$ and ensures that assisted tunneling terms are small.  The large confinement asymmetry enforces the hard-core constraint through a large on-site interaction, see \ref{sec:DDIEnergy}.  The coefficient $U$ can be tuned for ultracold polar molecules by an applied DC electric field.  One can determine the strength of the interaction $U=U^{\mathrm{DD},0}$ knowing only the expected dipole moment and the lattice constant from single-particle physics, and so scaling the phase diagram to $U$ is appropriate.

The iMPS method assumes that the many-body ground state has translational invariance under shifts by $q$ sites, and represents the wavefunction of the $q$-site unit cell as a matrix product state (MPS)~\cite{Schollwoeck_11} with entanglement cutoff $\chi$.  One minimizes the energy functional of the unit cell variationally in the parameters of the matrix product, using a sweeping procedure across sites in the unit cell reminiscent of the density-matrix renormalization group (DMRG) procedure for finite lattices~\cite{schollwock2005}.  Alternating with unit cell minimization is an updating of the effective environment of the unit cell using the most current parameters of the unit cell.  The procedure is halted when the 2-norm distance between the unit cell in successive optimizations drops below a desired tolerance.  Long-range interactions are facilitated within iMPS by using matrix product operators and fitting the interaction {$I_{j}$} to a sum of exponentials~\cite{Crosswhite_Doherty_08,Froewis_Nebendahl_10,Wall_Carr_13b}.  The advantage of the iMPS method for the problem at hand is its ability to handle strong long-range interactions without the significant boundary effects that plague DMRG-type procedures on finite lattices.  All results given here have been checked for convergence in the entanglement cutoff $\chi$ and the unit cell length $q$.

\begin{figure}[tbp]
\centerline{\includegraphics[width=1.0\columnwidth]{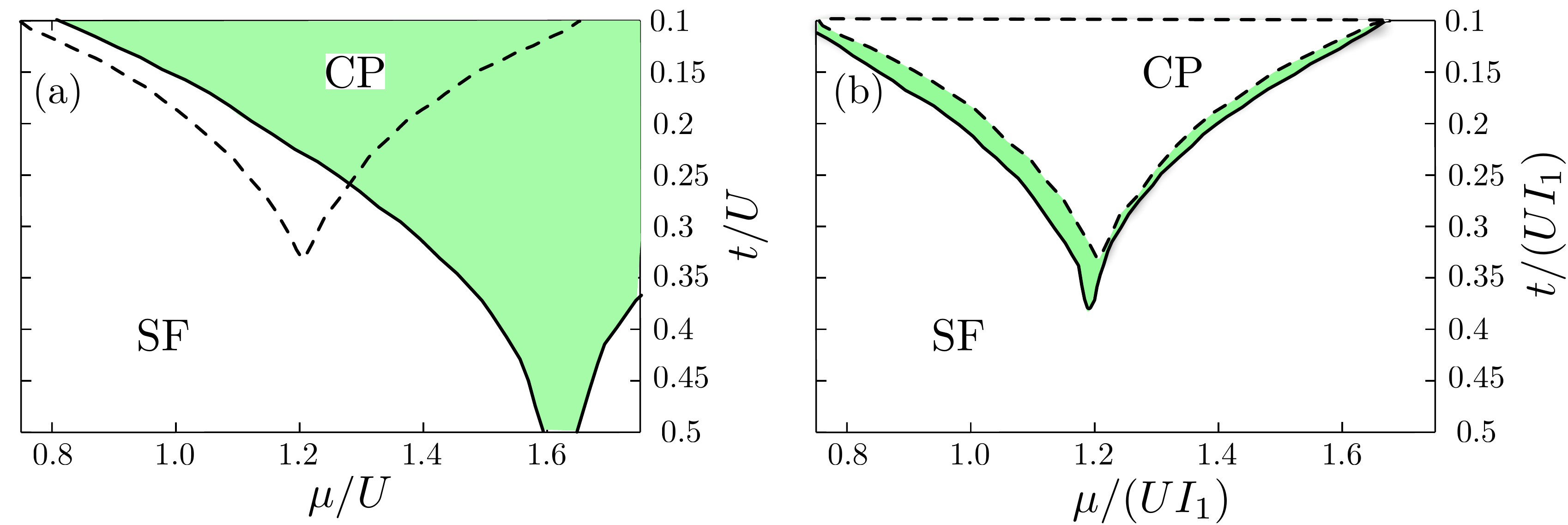}}
\caption{\label{fig:MB}  Bordered regions represent iMPS predictions of crystalline phase (CP).  Panel (a) displays the phase diagram with $U$ set to the unit of energy.  The region bounded by dashed lines uses the DFA, and has significant deviation from the green region with a solid boundary which uses the numerically determined effective dipole-dipole interaction.  Panel (b) is the same phase diagram with the unit of energy set to the nearest-neighbor interaction energy $UI_1$.  The effect of the confinement-modified interaction is not a simple rescaling of axes, as seen by the differences between the two bordered regions.}
\end{figure}

The phase diagrams of Eq.~\eref{eq:HCLRBH} for the DFA and lattice-modified dipolar interactions are shown in Fig.~\ref{fig:MB}.  The bordered areas represent predicted regions of crystalline phase (CP) with density $\rho=1/2$ and a non-vanishing single-particle gap.  The remainder of the plots are a gapless superfluid (SF) phase.  In the CP, the density correlation function $\mathcal{N}(r)\equiv \langle (\hat{n}_0-\rho)(\hat{n}_r-\rho)\rangle$ behaves as $\mathcal{N}(r)\to \mathrm{cst.}(-1)^r$ as $r\to \infty$ and the single-particle density matrix $\mathcal{A}(r)=\langle \hat{a}_0^{\dagger}\hat{a}_r\rangle$ is exponentially decaying with $r$.  In panel (a), the area bounded with dashed lines represents the region of CP computed using the DFA in the parameters $\mu/U$ and $\mu/U$.  The green area bordered with solid lines represents the CP boundaries computed with the actual lattice dipolar interaction.  The region predicted by the DFA is shifted both in chemical potential and tunneling by approximately {$I_1-1\approx 36\%$}.  However, the difference between the DFA and the true solution is not a simple rescaling of the axes.  This is shown in panel (b), which rescales the chemical potential and tunneling to the nearest-neighbor interaction energy, $\mu/(UI_{1})$ and {$t/(UI_1)$}.  Similar changes in the phase diagram will occur for the 2D case, with shifts of about 48\% for the parameters of Ref.~\cite{capogrosso2010}.  For soft-core particles which also possess local, isotropic interactions, the modified dipolar interaction will be relevant to the convexity of the interaction potential and hence to the formation of supersolid phases~\cite{Burnell}.

\section{Conclusions}
\label{sec:Concl}

In conclusion, we have shown that the dipole-dipole interaction is strongly modified by imperfect localization of particles in an unequally confined lattice, as found in many experiments in ultracold quantum gases.  In particular, we demonstrated that the effective dipole-dipole interactions in optical lattices, given by the matrix elements of the dipole-dipole interaction in a basis of lowest band Wannier functions, are significantly modified from the free-space form of the dipole-dipole interaction.  The modifications of the dipole-dipole interaction arise both from interactions in the exchange channel, controlled by the overlap between Wannier functions at different sites, and interactions in the direct channel.  The latter interactions depend crucially on the asymmetry in lattice confinement due to the $d$-wave symmetry of the dipole-dipole interaction.  We compared our results to results obtained by approximating Wannier functions as localized Gaussians.  Interactions in the direct channel are qualitatively reproduced by this approximation, though estimates of the magnitude can be off by significant factors.  Interactions in the exchange channel are quantitatively very poor in the exchange channel, underestimating interactions by orders of magnitude.  Based on numerical simulations, we put forward a simple characterization of the modified interaction as being exponential at moderate separations of a few lattice sites and power-law for large separations.  Using iMPS simulations, we showed that the modified interaction can significantly alter the predictions of many-body systems, including the determination of phase diagrams.

We acknowledge useful discussions with Mahadevan Ganesh, Kenji Maeda, and Zhigang Wu. This research was  supported in part by the National Science Foundation under Grants PHY-1207881 and NSF PHY11-25915, by AFOSR grant number FA9550-11-1-0224, the Heidelberg center for Quantum Dynamics, and by the Alexander von Humboldt Foundation.  We also acknowledge the Golden Energy Computing Organization at the Colorado School of Mines for the use of resources acquired with financial assistance from the National Science Foundation and the National Renewable Energy Laboratories.  We would like to thank the KITP for hospitality.

\appendix
\section{Numerical procedures to compute Hubbard parameters}
\label{sec:Numerical}
In this section, we discuss numerical techniques for computing geometrical integrals such as Eq.~\eref{eq:DDGeo} which determine the effective interaction between particles localized in lowest band Wannier states.  Such integrals may be written in the general form
\begin{eqnarray}
\label{eq:SMgeo} \mathcal{G}_{\mathbf{i_1i_2i_2'i_1'}}&=\int d\mathbf{r}\int d\mathbf{r}'f_{\mathbf{i}_1\mathbf{i}_1'}(\mathbf{r}) I\left(\mathbf{r}-\mathbf{r}'\right) f_{\mathbf{i}_2'\mathbf{i}_2'}(\mathbf{r}')\, ,
\end{eqnarray}
where $I\left(\mathbf{r}-\mathbf{r}'\right)$ is the dimensionless two-particle potential.  The first method employs the convolution theorem to express the integration over the primed coordinates in Eq.~\eref{eq:SMgeo} as a function of the unprimed coordinates with high accuracy.  The remaining three-dimensional integral is then performed with standard numerical quadrature.  We will call this method the real-space method.  In the second method, we compute the interaction matrix elements of the dipole-dipole interaction potential in the basis of lowest band Bloch functions.  The matrix elements in the Wannier basis are then obtained by quasimomentum Fourier transform.  We will refer to this latter method as the Bloch expansion method.  Both methods exhibit steep scaling with the linear domain size $L$ which prohibits studies of very large systems.  In order to ensure well-converged results, we restricted our analysis to a separation of at most 7 lattice sites, see Sec.~\ref{sec:Edd}.

\subsection{Real-space method}

In the real-space method, we begin with Eq.~\eref{eq:SMgeo} and apply the convolution theorem to find
\begin{eqnarray}
\label{eq:FinalFourierIntegral}&\mathcal{G}_{\mathbf{i_1i_2i_2'i_1'}}=\int d\mathbf{r} f_{\mathbf{i}_1\mathbf{i}_1'}(\mathbf{r})\mathcal{F}^{-1}_{\mathbf{r}}\{\mathcal{F}_{\mathbf{k}}[I\left(\mathbf{r}\right)]\mathcal{F}_{\mathbf{k}}[f_{\mathbf{i}_2\mathbf{i}_2'}(\mathbf{r}')]\}\, ,
\end{eqnarray}
where $\mathcal{F}_{\mathbf{k}}[g(\mathbf{r})]$ denotes the Fourier transform of the function $g(\mathbf{r})$ as a function of $\mathbf{k}$ and likewise for the inverse transform $\mathcal{F}^{-1}_{\mathbf{r}}\{\bullet\}$.  For example, the Fourier transform of the spatial part of the dipole-dipole potential is $\mathcal{F}_{\mathbf{k}}[{C^{(2)}_q(\mathbf{r})}/{r^3}]=- {4\pi}C^{(2)}_q(\mathbf{k})/3$,~\footnote{The Fourier transform of the dipole-dipole potential is ill-defined at $k=0$, but our results do not depend on this value.} and the Fourier transform of the delta-function potential is a constant.  Hence, the evaluation of the Hubbard parameter may be computed by three-dimensional Fourier transforms followed by a three-dimensional integration in real space rather than by a six-dimensional real space integral.  To perform these procedures numerically, we consider each Cartesian dimension to be a symmetric finite interval $S=[-L/2,L/2]$ with periodic boundary conditions, and discretize each interval with $n_g$ grid points.  The grid spacing in the discrete Fourier conjugate domain is $2\pi/L$ and the extent of the domain in Fourier space is controlled by $\pi n_g/L$, the inverse real space step size. The transformation from a function to its discrete Fourier conjugate is performed by the fast Fourier transform (FFT) algorithm in $\mathcal{O}(n_g^3\log n_g)$ time~\cite{Brigham_88}.  Because the Wannier functions on a finite domain are periodic and band-limited, their discrete and continuous Fourier transforms are related by a scaling constant provided we sample the entire domain at a frequency of at least twice their largest frequency component~\cite{Brigham_88}.  As is known for spectral methods, convergence of the Fourier space calculation in Eq.~\eref{eq:FinalFourierIntegral} is exponential in $L$ provided that $n_g/L$ is large enough to capture the full support of the function in Fourier space.  Defining $g$ to be the support of the lowest band Wannier function in the discrete Fourier space, the choice $n_g=2gL+1$ ensures that the function is fully captured in Fourier space.  The support of a Wannier function in Fourier space can be determined by using Parseval's theorem on finite Fourier subintervals to determine that the norm is unity to a desired tolerance.  For typical $g\sim 5-7$, the real space integration in Eq.~\eref{eq:FinalFourierIntegral} is of acceptable precision using a high-order Simpson integrator~\cite{press1993}.  

There are two dominant sources of error in the real space method.  The first error is due to the discretization of the real space domain and the associated discretization error of the numerical quadrature.  This error may be controlled by increasing $g$.  The second source of error is spurious interactions due to periodic boundary conditions.  While these interactions vanish as the domain becomes infinite, convergence may be slow due to, e.g.~,the power-law decay of the dipole-dipole interaction at long range.  Hence, we have instead used a finite-size scaling analysis to extrapolate our results to the limit of an infinite lattice.  The main limitation on the system sizes that we can reach using the real space method is a memory requirement which scales as $\mathcal{O}\left(g^3L^3\right)$ due to the non-separability of the dipole-dipole potential.  

\subsection{Bloch expansion method}

In the Bloch expansion method, we study the matrix elements of the interaction potential in the basis of lowest band Bloch functions $\psi_{\mathbf{q}}\left(\mathbf{r}\right)$,
\begin{eqnarray}
\label{eq:BO}\mathcal{V}_{\mathbf{q}_1\mathbf{q}_1'}^{\mathbf{q}_2\mathbf{q}_2'}&=\int d\mathbf{r}\int d\mathbf{r}' \left[\psi_{\mathbf{q}_1}\left(\mathbf{r}\right)\psi_{\mathbf{q}_2}\left(\mathbf{r}'\right)\right]^{\star}I\left(\mathbf{r}-\mathbf{r}'\right)\left[\psi_{\mathbf{q}_1'}\left(\mathbf{r}\right)\psi_{\mathbf{q}_2}\left(\mathbf{r}'\right)\right]\, .
\end{eqnarray}
The geometric integral Eq.~\eref{eq:SMgeo} is then given by
\begin{eqnarray}
\mathcal{G}_{\mathbf{i_1i_2i_2'i_1'}}&=\frac{1}{L^2}\sum_{\mathbf{q}_1\mathbf{q}_2\mathbf{q}_1'\mathbf{q}_2'}e^{i\mathbf{q}_1\cdot\mathbf{r}_{i_1}}e^{i\mathbf{q}_2\cdot\mathbf{r}_{i_2}}e^{-i\mathbf{q}_1'\cdot\mathbf{r}_{i_1'}}e^{-i\mathbf{q}_2'\cdot\mathbf{r}_{i_2'}}\mathcal{V}_{\mathbf{q}_1\mathbf{q}_1'}^{\mathbf{q}_2\mathbf{q}_2'}\, , 
\end{eqnarray}
see Eq.~(\ref{eq:wanndef}).  Using the fact that the interaction potential depends only on the relative coordinate, we find
\begin{eqnarray}
\mathcal{V}_{\mathbf{q}_1\mathbf{q}_1'}^{\mathbf{q}_2\mathbf{q}_2'}&=e^{i\mathbf{R}\cdot \mathbf{Q}}\mathcal{V}_{\mathbf{q}_1\mathbf{q}_1'}^{\mathbf{q}_2\mathbf{q}_2'}
\end{eqnarray}
where $\mathbf{R}$ is any Bravais lattice vector and $\mathbf{Q}\equiv \mathbf{q}_1'+\mathbf{q}_2'-\mathbf{q}_1-\mathbf{q}_2$ modulo $2\pi$.  This implies that there are only $N^3$ independent non-vanishing matrix elements in Eq.~\eref{eq:BO} for a lattice with $N$ unit cells, as opposed to the $N^4$ possible configurations of the four quasimomenta.   As in the main text, we will study the case of the simple cubic lattice in which the Bloch functions separate along principal axes,
\begin{eqnarray}
\psi_{\mathbf{q}}\left(\mathbf{r}\right)&=\prod_{\nu=x,y,z}\psi_{q_{\nu}}\left(r_{\nu}\right)\, .
\end{eqnarray}
The separability of the Bloch functions allows us to write the integral Eq.~\eref{eq:BO} in the form
\begin{eqnarray}
\fl \mathcal{V}_{\mathbf{q}_1\mathbf{q}_1'}^{\mathbf{q}_2\mathbf{q}_2'}&=&\prod_{\nu=x,y,z}\int_0^L d{r}_{\nu} \int_0^{L} d{r}_{\nu}' \left[\psi_{{q}_{1\nu}}\left({r}_{\nu}\right)\psi_{{q}_{2\nu}}\left({r}_{\nu}'\right)\right]^{\star}I\left(\mathbf{r}-\mathbf{r}'\right)\left[\psi_{{q}_{1\nu}'}\left({r}_{\nu}\right)\psi_{{q}_{2\nu}'}\left({r}_{\nu}'\right)\right]\, .
\end{eqnarray}
Changing integration variables to $2\xi_{\nu}\equiv r_{\nu}+r_{\nu}'$, $2\eta_{\nu}\equiv r_{\nu}-r_{\nu}'$ with Jacobian 2 along each Cartesian direction and expanding the one-dimensional Bloch functions as
\begin{eqnarray}
\psi_{q}\left(x\right)&=\lim_{\ell\to\infty} \frac{1}{\sqrt{L}}e^{iqx}\sum_{p=-\ell}^{\ell}c_q^pe^{2\pi i p x}\, ,
\end{eqnarray}
with $L$ the number of lattice sites along each Cartesian direction, we find
\begin{eqnarray}
\fl \mathcal{V}_{\mathbf{q}_1\mathbf{q}_1'}^{\mathbf{q}_2\mathbf{q}_2'}&=&\frac{2^3}{L^6}\prod_{\nu=x,y,z}\sum_{p_{1\nu}p_{2\nu}p_{1\nu}'p_{2\nu}'} c_{q_1}^{p_1}c_{q_2}^{p_2}c_{q_1'}^{p_1'}c_{q_2'}^{p_2'} \sum_{f_{\nu}=\left\{-1,1\right\}}f_{\nu} \\
&\times &\int_0^{f_{\nu}L/2} d{\eta }_{\nu} e^{2\pi i\eta_{\nu}\Delta_{\nu}}I\left(2\boldsymbol{\eta}\right) \int_{f_{\nu}\eta_{\nu}}^{L-f_{\nu}\eta_{\nu}} d{\xi}_{\nu} e^{2\pi i t_{\nu}\xi_{\nu}}\, .
\end{eqnarray}
Here, $\ell$ is a finite Fourier cutoff used in numerics and we have defined
\begin{eqnarray}
t_{\nu}&\equiv p_{1\nu}'-p_{1\nu}+p_{2\nu}'-p_{2\nu}+\frac{q_{1\nu}'-q_{1\nu}+q_{2\nu}'-q_{2\nu}}{2\pi}\,, \\
\Delta_{\nu}&\equiv p_{1\nu}'-p_{1\nu}+p_{2\nu}-p_{2\nu}'+\frac{q_{1\nu}'-q_{1\nu}+q_{2\nu}-q_{2\nu}'}{2\pi}\,.
\end{eqnarray}
The integrals over $\xi_{\nu}$ are
\begin{eqnarray}
\int_{f_{\nu}\eta_{\nu}}^{L-f_{\nu}\eta_{\nu}} d{\xi}_{\nu} e^{2\pi i t_{\nu}\xi_{\nu}}&=L\delta_{t_{\nu},0}-\frac{\sin\left(2\pi \eta_{\nu} t_{\nu}\right)}{\pi t_{\nu}}\, .
\end{eqnarray}
We will keep only the term proportional to the Kronecker delta, as this is the dominant contribution for large lattices.  This approximation becomes exact in the limit of an infinite lattice, $L\to \infty$, as has been shown for the delta-function potential in Ref.~\cite{Wall_Carr_13}.  For simplicity, we now also require that the interaction potential is invariant under inversion by any Cartesian coordinate.  This is true for the $q=0$ component of the dipole-dipole interaction, as well as for the delta function potential.  However, the method directly extends to more general interactions.  With these two conditions, we find in the limit as $L\to \infty$,
\begin{eqnarray}
\nonumber \fl\mathcal{V}_{\mathbf{q}_1\mathbf{q}_1'}^{\mathbf{q}_2\mathbf{q}_2'}&=&\frac{1}{L^3}\prod_{\nu=x,y,z}\sum_{p_{1\nu}p_{2\nu}p_{1\nu}'p_{2\nu}'} \delta\left(p_{1\nu}'-p_{1\nu}+p_{2\nu}'-p_{2\nu}+\frac{q_{1\nu}'-q_{1\nu}+q_{2\nu}'-q_{2\nu}}{2\pi}\right)\\
&\times &c_{q_1}^{p_1}c_{q_2}^{p_2}c_{q_1'}^{p_1'}c_{q_2'}^{p_2'}\mathcal{F}_{\pi \boldsymbol{\Delta}}\left[I\right]\, ,
\end{eqnarray}
where $\mathcal{F}_{\mathbf{k}}\left[I\right]$ is again the Fourier transform of $I$ as a function of $\mathbf{k}$ and $\boldsymbol{\Delta}=\left(\Delta_x,\Delta_y,\Delta_z\right)$.

The Bloch expansion method has the advantage of not introducing any discretization error, and also requires significantly less memory when the lattice is separable.  However, the computational scaling of this method is $\mathcal{O}\left(L^9\right)$ as opposed to $\mathcal{O}\left(L^3\right)$ for the real space method.  The Bloch expansion method suffers from spurious interactions due to periodic boundary conditions, just as in the real space method.  We find that both methods agree when we extrapolate to the limit $L\to \infty$, and allow us to put a conservative bound of 1\% on our estimated error.  Additionally, we have benchmarked both methods with the case of a delta-function potential, where both analytical and numerically exact results are available.

\section{Dipole-Dipole Interaction energy of the anisotropic harmonic oscillator}
\label{sec:DDIEnergy}
In this appendix we derive the dipolar interaction energy of two particles in the ground state of an anisotropic harmonic oscillator, Eq.~(\ref{eq:U0ho}) of the main text, and compare the results with the on-site interaction energy in an optical lattice.  As in the main text, we use the quasi-2D geometry $\ell_x=\ell_y\equiv\ell_{\perp}$ and choose the harmonic oscillator lengths to match the local curvature of a lattice site minimum via
\begin{eqnarray}
\label{eq:hoconsistency}\ell_{\nu}=\frac{a}{\pi\tilde{V}_{\nu}^{1/4}}\, .
\end{eqnarray}
The ground state wave function may be written in cylindrical coordinates as
\begin{eqnarray}
\psi\left(\rho,\phi,z\right)&=\frac{1}{\ell_{\perp}\sqrt{\ell_z}\pi^{3/4}}\exp\left(-\frac{z^2}{2\ell_z^2}-\frac{\rho^2}{2\ell_{\perp}^2}\right)\, ,
\end{eqnarray}
where $\rho^2=x^2+y^2$.  Using the convolution theorem, we can write the {dimensionless }dipolar interaction energy as{
\begin{eqnarray}
\label{eq:UIE} I_{0}^{\mathrm{ho}}&=\frac{a^3}{2\pi^2}\int d\mathbf{k}\left(\cos^2\theta-1/3\right)n^2\left(\mathbf{k}\right)\, ,
\end{eqnarray}
}where
\begin{eqnarray}
n\left(\mathbf{k}\right)=n\left(\rho,z\right)&=\exp\left(-\frac{z^2\ell_z^2}{4}-\frac{\rho^2\ell_{\perp}^2}{4}\right)
\end{eqnarray}
is the Fourier transform of the density and $\theta$ is the angle between $\mathbf{k}$ and the $z$-axis.  Performing the integration in Eq.~\eref{eq:UIE} over $z$ and $\phi$ yields{
\begin{eqnarray}
\fl I_{0}^{\mathrm{ho}}&=&a^3\int_{0}^{\infty}d\rho \rho \exp\left(-\frac{\rho^2\ell_{\perp}^2}{2}\right)\Big\{\frac{1}{3}\sqrt{\frac{8}{\pi}}\frac{1}{\ell_z}-\rho\exp\left(\frac{\rho^2\ell_z^2}{2}\right)[1-\mathrm{erf}\left(\frac{\rho\ell_z}{\sqrt{2}}\right)]\Big\}\, ,
\end{eqnarray}
}with $\mathrm{erf}\left(x\right)$ the error function.  Integrating over $\rho$ yields{
\begin{eqnarray}
\label{eq:Uho}\fl I_{0}^{\mathrm{ho}}&=&\frac{a^3}{3}\sqrt{\frac{8}{\pi}}\frac{1}{\ell_z\ell_{\perp}^2}+\sqrt{\frac{2}{\pi}}\frac{\ell_z}{\ell_{\perp}^2}\frac{1}{\ell_{\perp}^2-\ell_z^2}-\sqrt{\frac{2}{\pi}}\frac{1}{\left(\ell_{\perp}^2-\ell_z^2\right)^{3/2}}\cot^{-1}\left(\frac{\ell_z}{\sqrt{\ell_{\perp}^2-\ell_z^2}}\right)\, .
\end{eqnarray}
}Inserting the definitions from the main text, {$\bar{\ell}\equiv (\ell_z\ell_{\perp}^2)^{1/3}/a$}, $\alpha\equiv \ell_z/\ell_{\perp}$, and $\beta\equiv \alpha (1-\alpha^2)^{-1/2}$, we find
\begin{eqnarray}
\label{eq:U0hos} I_0^{\mathrm{ho}}&= \frac{\sqrt{2}}{\bar{\ell}\,^3\pi}[\frac{2}{3}+\beta^2-{\beta}(1-\alpha^2)^{-1}\cot^{-1}(\beta)]\, ,
\end{eqnarray}

\begin{figure}[ht]
\centerline{\includegraphics[width=0.5\columnwidth]{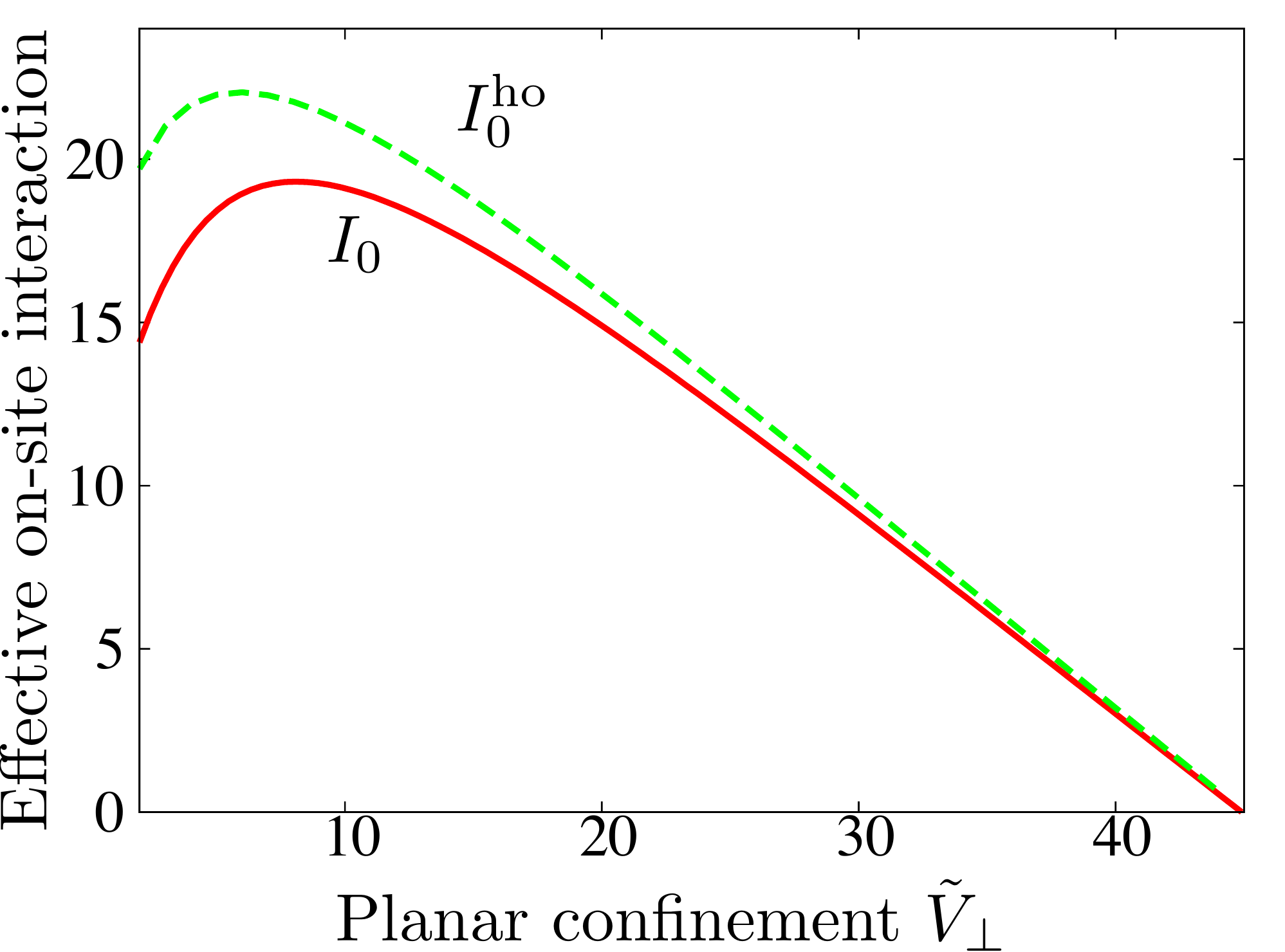}}
\caption{\label{fig:Uho}  The on-site dipolar interactions in a quasi-2D lattice, $I_{0}$ (red solid line), and the dipolar interaction energy of a cylindrically symmetric harmonic oscillator with the same local curvature as the lattice site, $I_0^{\mathrm{ho}}$ (green dashed line), have similar qualitative behavior with respect to confinement asymmetry.  In particular, both vanish as the confinement becomes isotropic, $\tilde{V}_{\perp}=\tilde{V}_{z}=45$, and are significant for large confinement asymmetry.}
\vspace{-0.2in}
\end{figure}

A comparison between the function Eq.~\eref{eq:U0hos} with the oscillator lengths chosen as in Eq.~\eref{eq:hoconsistency} and the result computed via Wannier functions as explained in \ref{sec:Numerical} is given in Fig.~\ref{fig:Uho}.  We find that the on-site interaction energy in the harmonic oscillator is always larger than the corresponding energy in an optical lattice, often by 10\%.  Additionally, the on-site interaction is large for small to moderate $V_{\parallel}$, enforcing the hard-core constraint used in the many-body study in the main text.

\bibliographystyle{prsty}
\bibliography{running}

\end{document}